\documentclass[aps,prl,a4paper,twocolumn]{revtex4}

\usepackage{graphicx}
\usepackage{ae,aecompl}
\usepackage{bm}
\usepackage{amsmath}
\usepackage{textcomp}

\newcommand{\kB}{k_{\rm B}}
\newcommand{\cosphi}{\langle \cos \phi \rangle}

\advance\textheight by 0.8cm
\advance\textwidth by 0.15cm
\advance\hoffset by -0.1cm

\begin{document}

\title{Squeezing and entanglement in a Bose-Einstein condensate}
\author{J. Est\`eve}
\author{C. Gross}
\author{A. Weller}
\author{S. Giovanazzi}
\author{M. K. Oberthaler}
\email{entanglement@matterwave.de}
\affiliation{Kirchhoff Institut f\"ur Physik, Universit\"at Heidelberg, Im Neuenheimer Feld 227, 69120 Heidelberg, Germany}

\maketitle

\textbf{Entanglement, a key feature of quantum mechanics, is a resource that
allows the improvement of precision measurements beyond the
conventional bound reachable by classical
means~\cite{giovannetti:2004ix}. This is known as the standard
quantum limit, already defining the accuracy of the best available
sensors for various quantities such as time~\cite{santarelli:1999aa}
or position~\cite{goda:2008ix,arcizet:2006eu}.
Many of these sensors are interferometers in which the standard quantum
limit can be overcome by feeding their two input ports with
quantum-entangled states, in particular spin squeezed
states~\cite{kitagawa:1993wd,wineland:1994rr}. For atomic
interferometers, Bose-Einstein condensates of ultracold atoms are
considered good candidates to provide such states involving
a large number of particles~\cite{sorensen:2001wd}. In this
letter, we demonstrate their experimental realization by splitting a
condensate in a few parts using a lattice potential. Site
resolved detection of the atoms allows the measurement of the
conjugated variables atom number difference and relative phase. The
observed fluctuations imply entanglement between the
particles~\cite{sorensen:2001wd,wang:2003eu,korbicz:2005oq}, a resource that would allow a precision gain of 3.8~dB over the standard quantum limit for interferometric measurements.}

\begin{figure*}
\begin{center}
\includegraphics{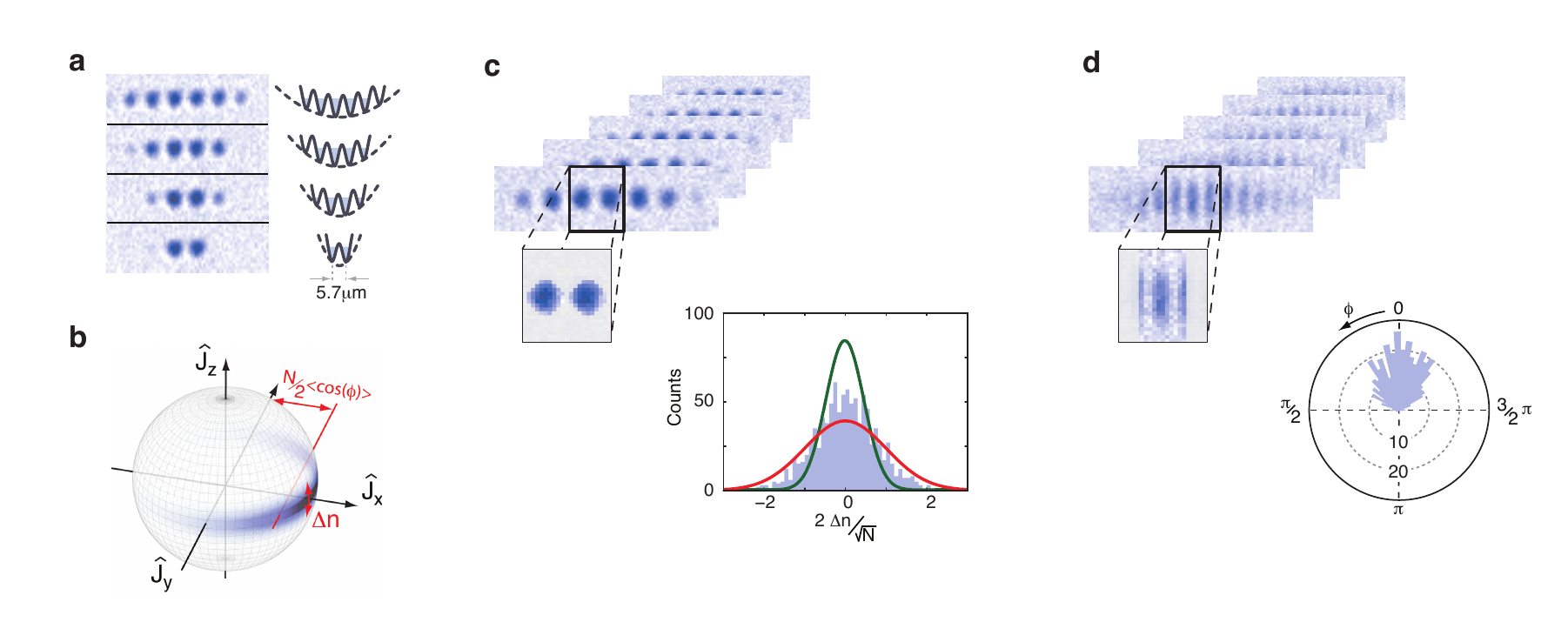}
\caption{\textbf{Observing spin squeezing in a Bose-Einstein condensate confined in
a double- or six-well trap}. \textbf{a)} The atoms are trapped in an
optical lattice potential superimposed on an harmonic dipole trap.
The number of occupied sites is adjusted by changing the confinement in lattice direction. High-resolution imaging allows us to resolve each site. \textbf{b)} Gain in quantum metrology is obtained for
spin squeezed states exhibiting reduced fluctuations in one
direction $(z)$ and a large enough polarization in the orthogonal
plane $(x,y)$ as depicted on the Bloch sphere. For our system, spin
fluctuations in the $z$ direction translate to atom number
difference fluctuations $\Delta n$ between two adjacent wells. The
polarization of the spin in the $x-y$ plane is proportional to the
phase coherence, $\cosphi$, between the wells. \textbf{c)} The atom
number fluctuations at each site are measured by integrating the
atomic density obtained from absorption images. A typical histogram
showing sub-Poissonian fluctuations in the atom number difference is compared with the binomial distribution (red curve). The green curve corresponds to the deduced distribution after subtracting the photon shot noise, leading to a number squeezing factor of $\xi_N^2=-6.6$~dB. \textbf{d)} The phase
coherence is inferred from the interference patterns between adjacent wells. The shown histogram corresponds to a phase coherence of $\cosphi = 0.9$.}
\label{fig.setup}
\end{center}
\end{figure*}

Spin squeezing was one of the first quantum strategies proposed to
overcome the standard quantum limit in a precision
measurement~\cite{kitagawa:1993wd,wineland:1994rr} that triggered
many experiments~\cite{hald:1999lh,kuzmich:2000bv,julsgaard:2001oz,chaudhury:2007le,fernholz:2008ez,meyer:2001hc,leibfried:2004hs,roos:2004dp}.
It applies to measurements where the final readout is done by
counting the occupancy difference between two quantum states, as in
interferometry or in spectroscopy. The name "spin squeezing"
originates from the fact that the $N$ particles used in the
measurement can be described by a fictitious spin $J=N/2$. In an
interferometric sequence, the spin undergoes a series of rotations
where one of the rotation angles is the phase shift to be measured.
A sufficient criterion for the input state allowing for quantum
enhanced metrology is given by $\xi_S<1$ where $\xi^2_S = 2 J \Delta
J^2_z/(\langle J_x\rangle^2 + \langle J_y\rangle^2)$ is the
squeezing parameter introduced in ref.~\cite{wineland:1994rr}. The fluctuations of the spin in one direction have to be reduced
below shot-noise $\Delta J^2_z < J/2$, and the spin
polarization in the orthogonal plane $\langle J_x \rangle^2 +
\langle J_y \rangle^2$ has to be large enough to maintain the
sensitivity of the interferometer. A pictorial representation of
this condition is shown in figure~\ref{fig.setup}b. The precision of
such a quantum enhanced measurement is $\xi_S/\sqrt{N}$, whereas
the standard quantum limit set by shot-noise is $1/\sqrt{N}$.

In this Letter, we report on the observation of entangled squeezed
states in a Bose-Einstein condensate of $^{87}$Rb atoms. The
particles are distributed over a small number of lattice sites
(between 2 and 6) in a one dimensional optical lattice (see
figure~\ref{fig.setup}a). The occupation number per site ranges from
100 to 1100 atoms. The two modes supporting the squeezing are two
states of the external atomic motion corresponding to the condensate
mean-field wave-functions in two adjacent lattice sites. These modes
are spatially well separated and thus represent an ideal starting
condition for a spatially split interferometer. Labeling $a^\dagger$
and $b^\dagger$ the creation operators associated with the two
modes, the fictitious spin components can be defined as $J_x =
(a^\dagger b + b^\dagger a)/2$, $J_y = i(a^\dagger b - b^\dagger
a)/2$ and $J_z=(a^\dagger a - b^\dagger b)/2$. The $z$ component
corresponds to half the atom number difference between the wells.
Because the mean occupation numbers in the two wells, $n_a$ and $n_b$, are large, the expectation value of the $x,y$ components can be
approximated by $\langle J_x \rangle \simeq \sqrt{n_a \, n_b}
\cosphi$, $\langle J_y \rangle \simeq \sqrt{n_a \, n_b} \langle \sin
\phi \rangle$, respectively,  where $\phi$ is the phase difference between the two macroscopic wave-functions.

Spin squeezing by means of unitary evolution requires a non-linear component
in the Hamiltonian~\cite{kitagawa:1993wd}; this is provided by the
repulsive interactions between the atoms of the condensate. The
corresponding suppression of atom number fluctuations in a
Bose-Einstein condensate has been indirectly
observed~\cite{orzel:2001aa,greiner:2002aa,gerbier:2006cr,sebby-strabley:2007nx,jo:2007hl,li:2007jx}.
However, by definition of the squeezing factor $\xi_S$, its
experimental determination requires to access the local properties
of the atoms occupying the two sites of interest. By imaging the
condensate with a resolution of 1~$\mu$m (fwhm), which is well below the
lattice spacing of 5.7~$\mu$m, we fulfill this criterion of a local
measurement. The wells of the lattice are fully resolved (see
figure~\ref{fig.setup}a), which allows the determination of the atom
number in each lattice site by direct integration of the atomic
density as obtained by absorption imaging. Local interference
measurements after a short time of expansion of the condensate such
that only neighboring sites overlap reveal the phase between these
wells. In Figures~\ref{fig.setup}c and d, we display typical data sets
for the two types of measurement. The technical details for the precise experimental procedures and calculations used to deduce the number squeezing factor and the phase coherence are given in the supplementary information. The fluctuation measurement of the two
conjugated variables, number and phase, yields information about the
quantum state of the system and, in particular, allows the detection of  macroscopic entanglement between the particles.

\begin{figure}
\begin{center}
\includegraphics{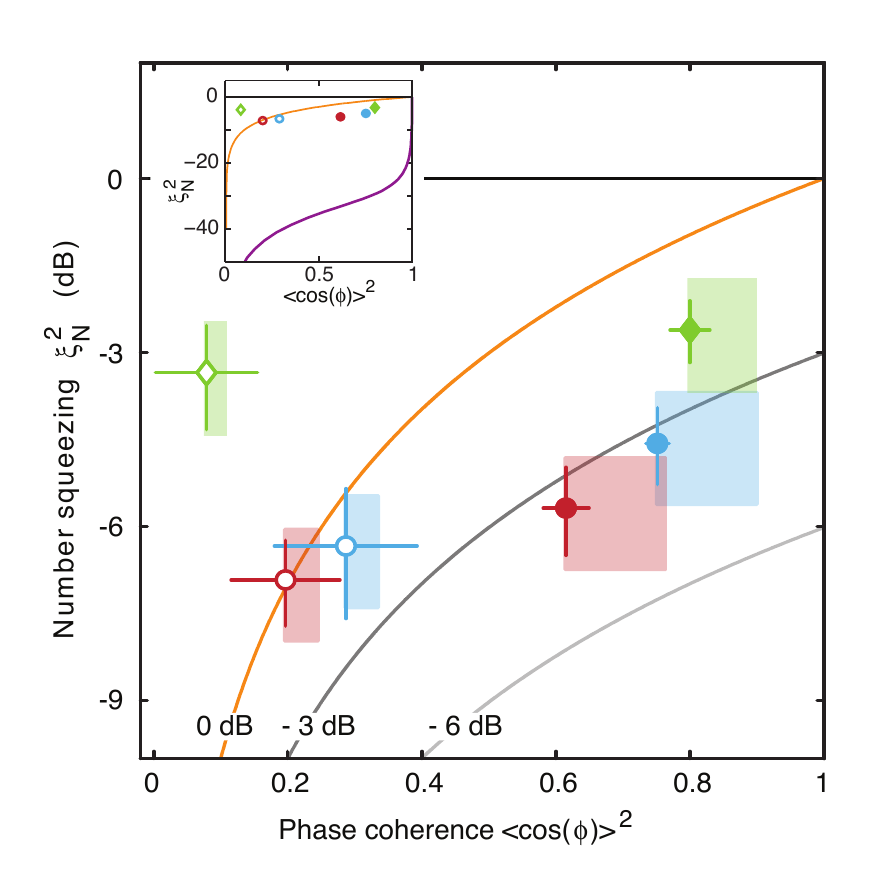}
\caption{\textbf{Number squeezing and phase coherence as a criterion for quantum metrology and entanglement.} Measurements are shown for the two main well pairs of a six-well lattice (red and blue circles) and for a double-well potential (green diamonds). The total atom number, $N$, in each pair is approximately 2200 in the six-well case and 1600 in the double-well case. Filled and open symbols discriminate two different experimental regimes corresponding to different final lattice depths (see figure~\ref{fig.squeezing}). For filled symbols, number squeezing and high phase coherence are simultaneously observed, whereas for the open symbols the phase coherence is degraded. The limits for different quantum metrology gains $\xi_S^2$ are plotted. In the coherent-number-squeezed region (below the orange line), directly usable entanglement is necessarily present in the system. The shaded areas show systematic error bounds due to a possible miscalibration of the atom number ($\pm$20\%) and to an underestimation of the phase coherence caused by technical noise. The error bars indicate standard deviation deduced from at least 400 experimental realizations. The inset (same quantities as main panel) shows that our data are approximately 25~dB above the optimal allowed number squeezing (purple line), showing room for improvement~\cite{sorensen:2001ax}.}
\label{fig.entanglement}
\end{center}
\end{figure}

\begin{figure*}
\begin{center}
\includegraphics{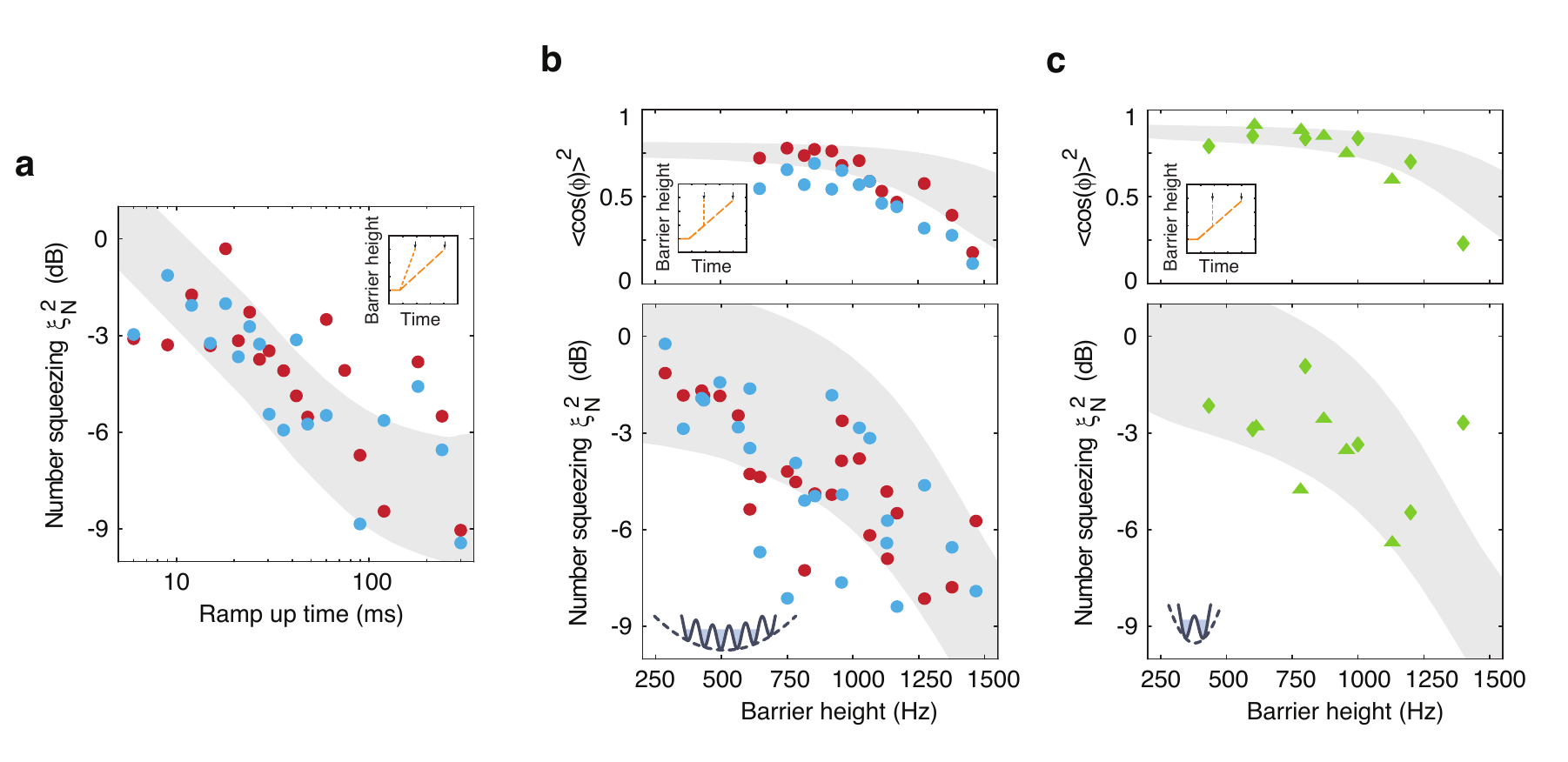}
\caption{\textbf{Systematics on number squeezing and phase coherence during
the splitting of a condensate.} \textbf{a)} Investigating the
adiabaticity condition for the evolution of the many-body state
during the splitting process. The number squeezing factors for the
two most populated well pairs (blue and red circles) are measured
after ramping up the lattice depth from the value at which the condensate is obtained (430~Hz) to a fixed end value (1650~Hz), for a number of different total ramp times (see inset). Numerical simulations using the two-mode Josephson model reproduce the general observed behavior, assuming an initial thermal population
of the density matrix corresponding to temperatures between $20$~nK
and $40$~nK (grey shaded area).  \textbf{b)} Number squeezing and phase coherence for
different final lattice depths. The ramp speed is fixed to 4~Hz.ms$^{-1}$
(300~ms ramp time in a) to satisfy the adiabaticity criterion. Number squeezing improves before the phase coherence drops, leading to coherent number squeezing. The grey shaded area shows the predictions of the two-mode model assuming an adiabatic evolution of the density matrix with initial temperatures between $10$~nK and $30$~nK (three to ten populated many-body states).
\textbf{c)} Same as b) for a double-well. The ramp speed is 2~Hz/ms and 8~Hz/ms for the triangles and diamonds respectively. The symbols shown in figure~\ref{fig.entanglement} correspond to the average of all the data measured for barrier heights between 650 and 900~Hz (filled symbols) and above 1300~Hz (open symbols) in the six-well case (b) and between 650 and 1200~Hz (filled symbols)  and above 1400 ~Hz (open symbols) in the double-well case (c). Some data measured at very high lattice depth in the double-well case that participate in the averaging do not appear here.}
\label{fig.squeezing}
\end{center}
\end{figure*}

Figure~\ref{fig.entanglement} summarizes the conjugated number -
phase measurements in different experimental situations. The
vertical axis corresponds to the number squeezing parameter,
$\xi^2_N=\Delta J_z^2 / \Delta J_{z,{\rm ref}}^2$ which measures how
much the fluctuations of the atom number difference are suppressed
in comparison with a binomial distribution with variance $\Delta J_{z,{\rm
ref}}^2 = n_a n_b /N$, which is expected in a non-squeezed situation. The
phase coherence, $\cosphi$, between the two wells defines the
horizontal axis. We choose the origin of the phase such that the
$J_y$ component has a zero mean value. In this case, the relevant squeezing parameter for quantum metrology, $\xi_S$, is given by $\xi_S = \xi_N /
\cosphi$. In the following, we will refer to it as the coherent number squeezing  parameter. Lines corresponding to $\xi_S^2 = -3$~dB and -6~dB
are plotted in the figure. The different data points correspond to different preparations of the atoms in the lattice, as detailed in the caption of figure~\ref{fig.squeezing}, and to different numbers of occupied sites. The simultaneous observation of number squeezing and high phase
coherence (solid symbols) reveals the presence of coherent number squeezing. In the double-well and six-well situation, we deduce best squeezing factors $\xi_S^2=-2.3$~dB and $\xi_S^2=-3.8$~dB, respectively.
The statistical error bars and bounds for systematic errors are indicated in the figure. 
The open symbols show that atom number fluctuations can be further suppressed at the expense of lower phase coherence. In the six-well situation, we observe number squeezing down to $\xi_N^2=-7.2$~dB,  which corresponds to fluctuations of 15 atoms per well, out of 1100.
The inset in figure~\ref{fig.entanglement} shows the optimal number squeezing for a given phase coherence~\cite{sorensen:2001ax} in our
experimental situation of 2200 atoms in a double-well, revealing
that there is still a great potential for improvement. Our best
measurements yields number fluctuations approximately 25~dB higher 
than the Heisenberg limited states with the same phase coherence.

Entanglement in the context of spin squeezing has been intensively discussed~\cite{sorensen:2001wd,wang:2003eu,korbicz:2005oq}. In a first quantization approach, it can be defined as the non separability of the $N$ body density matrix. With this definition, a sufficient criterion for entanglement coincides with the criterion for quantum metrology, namely $\xi_S<1$, which identifies spin-squeezing type entanglement as a useful resource~\cite{sorensen:2001wd}. In the context of indistinguishable bosons, as in our experiment, the number squeezing $\xi_N<1$ has been shown to be a sufficient criterion for the nonseparability of the reduced two-body density matrix~\cite{wang:2003eu,korbicz:2005oq}. This criterion is fulfilled for all the measurements shown in figure~\ref{fig.entanglement}. However, without the possibility of accessing the single-particle spin properties (by contrast to the case of an ion string~\cite{korbicz:2006kb}), this type of bipartite entanglement may not be usable as a resource. For this reason, only measurements in the coherent number squeezed region (solid symbols) indicate the definite presence of useful entanglement.

To identify what limits the amount of squeezing, we
consider the two-mode Josephson Hamiltonian $E_C/2 J_z^2 - 2E_J/N
J_x$ that describes two weakly coupled condensates. The Josephson energy, $E_J$, and the charging energy, $E_C$, respectively characterize the tunneling rate between the two condensates and the repulsive interaction energy inside each well. Because the depletion of the condensates in each well is small ($\leq$12 atoms) we neglect intra-well excitation. Longer wavelength excitations that exist in the many well situation are also not considered. For the purpose of identifying the limiting factors on squeezing, this simplified model is sufficient because it captures the correct scaling of next neighbor fluctuations with temperature, tunneling rate and interaction energy~\cite{javanainen:1999ab}.
In the two-mode model and at temperatures $T$ high compared to the plasma
energy, $\sqrt{E_C E_J}/\kB$, the thermal excitation of the Josephson plasma mode limits the number squeezing to $\xi_N^2 \sim \kB T/\mu$ where $\mu$ is the chemical potential that measures the strength of the interaction $(E_C \sim \mu/N)$. In a typical experimental situation at
thermal equilibrium, these fluctuations are a strong limit on
both number and coherent number squeezing, as it is difficult to evaporatively cool the condensate much below the chemical potential.

To circumvent this limitation, we prepare the condensate in
a shallow lattice before increasing the lattice depth. During this
splitting process, the tunneling rate $E_J$ decreases from its
initial value, $E_{J}^{(i)}$, to a final value, $E_{J}^{(f)}$, while
the interaction energy $E_C$ stays almost constant in our
experimental situation. For sufficiently slow ramp speeds, the evolution
is expected to be adiabatic. Because the energy of the collective Josephson
mode decreases during the splitting, energy is removed
from the system, leading to an effective cooling of the relevant
degree of freedom. 
In the regime where only the linear part of the collective spectrum is populated, the effective temperature evolves as $T_{\rm eff}=T\sqrt{E_J^{(f)}/E_J^{(i)}}$ leading to a final number squeezing $\xi_{N,{\rm f}}^2 \approx \xi_{N,{\rm i}}^2 ( T_{\rm eff} / T)$, where $ \xi_{N,{\rm i}}^2 = 4 \kB T/(N
E_C)$ is the initial squeezing at equilibrium with temperature $T$. The optimal number squeezing, assuming adiabatic evolution, is obtained by reducing further the Josephson energy to enter the Fock regime $(E_J^{(f)}<E_C)$ where the collective spectrum is quadratic. The final number squeezing is then given by $\xi_{N,{\rm f}}^2 \simeq \xi_{N,{\rm i}}^2 \, (\kB T / 2 E_J^{(i)})$.

The assumption of adiabaticity is investigated by performing a
series of experiments in the six-well situation. The lattice depth is
ramped from the initial value at which the condensate is obtained
(430~Hz) to a fixed end value (1650~Hz) in different times. In
figure~\ref{fig.squeezing}a, the final number squeezing factors for
the two most populated well pairs are plotted versus the total
ramping time. As expected, no improvement in the number squeezing is observed for too fast ramps ($<20$~ms). For slower ramps, the number squeezing increases and saturates at $\xi_{N,{\rm f}}^2 = -6.6^{+0.8+0.8}_{-1.0-0.8}$~dB, where the given uncertainties are one-sigma statistical errors obtained
from approximately 1000 experimental realizations followed by systematic error bounds (see supplementary information). This improvement
in number squeezing by approximately a factor of approximately three is a
consequence of the effective cooling discussed above. As shown in
figure~\ref{fig.squeezing}a, the observed behavior is reproduced by
a numerical simulation of the splitting process with the two-mode Josephson
Hamiltonian.

High phase coherence is the second ingredient for a coherent number squeezing factor $\xi_S<1$. This requires that the effective temperature of the Josephson mode is below the tunneling coupling $E_J$, which is the case for the initial lattice depth~\cite{gati:2006ab}. Increasing the lattice depth decreases the phase coherence. Best coherent number squeezing is obtained by adiabatically increasing the lattice depth to an optimal value such that coherence is still high and number squeezing has occured.
In order to find this optimum, we investigate the
number squeezing and the phase coherence as a function of tunneling
coupling. We linearly ramp up the barrier height to different end
values, keeping the ramp speed in the adiabatic regime.
Figure~\ref{fig.squeezing} presents the results obtained both for six (Fig.~3b)
and two (Fig.~3c) occupied wells.
In the six-well case, we identify an optimum barrier height range between 650 and 900~Hz, where we deduce a best coherent number squeezing $\xi_S^2
=-3.8^{+0.3+0.8}_{-0.4-1.8}$~dB averaging over all the measurements. In the double-well situation, averaging all the points between 650 and 1200~Hz, we obtain $\xi_S^2=-2.3^{+0.2+0.8}_{-0.6-1.5}$~dB. The deduced mean phase coherence and number squeezing in the optimal regions correspond to the solid data points shown in figure~\ref{fig.entanglement}.

\begin{figure}
\begin{center}
\includegraphics{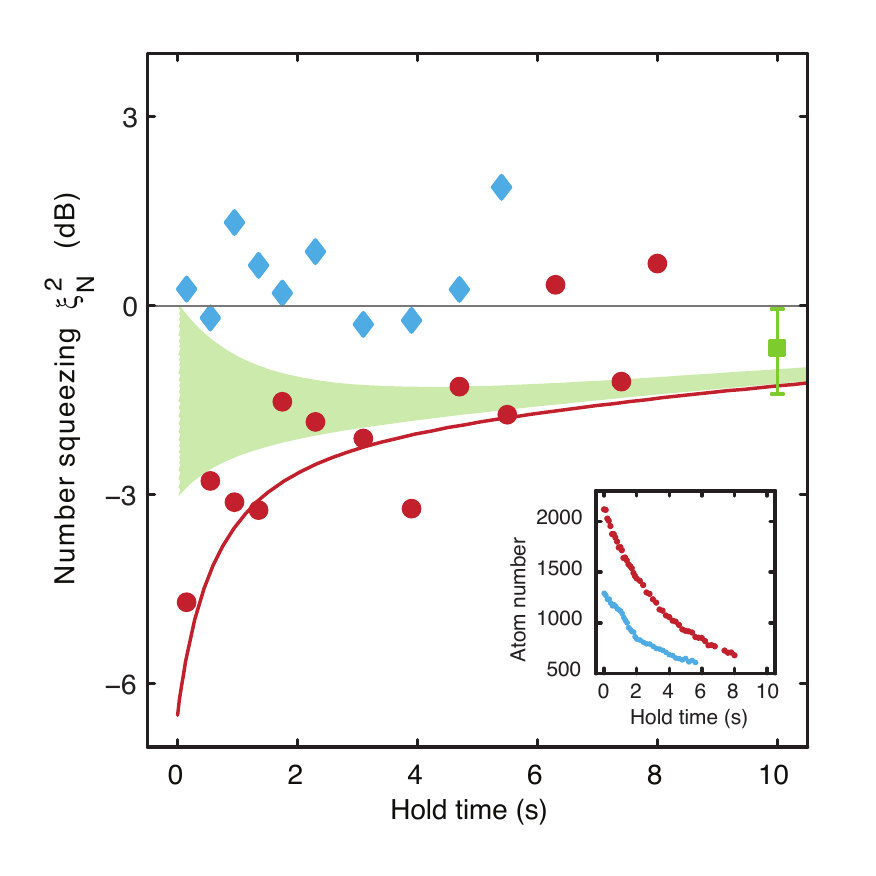}
\caption{\textbf{Random loss restores poissonian fluctuations.} Holding the
atoms in the lattice (six occupied sites), the atom number in the
decoupled two main well pairs decreases as shown in the inset, owing to
one- and three-body loss. Red circles show the evolution of number squeezing starting from a squeezed situation. The solid line is the prediction corresponding to the measured loss rates and initial squeezing. Blue diamonds show that in the absence of initial squeezing no further evolution is observed. The consistent behavior of the squeezing with loss is an independent check of the validity of our squeezing measurement technique. A more quantitative test is performed by holding the atoms 10~s and measuring the number squeezing inferred from 1000 realizations. The initial state is slightly squeezed ($-3 \, {\rm dB}< \xi_N^2 < 0\, {\rm dB}$) and a squeezing factor $-1.2 \, {\rm dB}< \xi_N^2< -1\, {\rm dB}$ (green shaded area) is expected at the time of measurement, when two-thirds of the atoms have been lost. We measure $\xi_N^2=-0.7^{+0.7}_{-0.7}$~dB consistent with the expected value within the $95\%$ statistical confidence bounds indicated by the error bar. The small improvement of number squeezing in the upper boundary of the green shaded area arises from the non-linear dependence of three-body loss on atom number. After a sufficient time, the one-body loss dominates and Poissonian fluctuations are restored.} \label{fig.loss}
\end{center}
\end{figure}

For the best observed number squeezing, we measure atom number fluctuations just above the detection threshold of our
absorption imaging technique. To reach this sensitivity
level, special care has to be taken to calibrate the deduced atom
number~\cite{reinaudi:2007nx} (see supplementary information). Furthermore, the contribution of the photon shot-noise has to be precisely measured and
subtracted~\cite{esteve:2005ab}. As an independent check of the
reliability of the atom counting, we monitor the evolution of number
squeezing with atom loss. It is well known in quantum optics that random
loss processes rapidly degrade the number squeezing. Red circles in figure~\ref{fig.loss} show this restoring effect with
a rate which is compatible with the measured one- and three-body loss rates. A further check is performed by monitoring the evolution of the number fluctuations for an initial state with almost Poissonian fluctuations; such a state is prepared by directly condensing the atoms in a deep lattice.  No change in the number squeezing is observed through the loss process, within the statistical errors.  During the measurement, the relevant parameters for the imaging, such as the extension of the cloud and its optical density, are kept constant.

To confirm the successful implementation of a quantitative atom number fluctuation measurement, we infer number squeezing
out of 1000 measurements after a 10~s holding time, starting from a
slightly squeezed situation ($-3 \, {\rm dB}< \xi_N^2 < 0\, {\rm dB}$).  At this time, two-thirds of the atoms are lost and a precise number squeezing factor $-1.2 \, {\rm dB}< \xi_N^2 < -1\, {\rm dB}$ is predicted from the measured one- and three- body loss coefficients. This exemplifies how particle losses can be used to prepare a well defined number distribution in a condensate. We measure $\xi_N^2=-0.7^{+0.7}_{-0.7}$~dB, where the indicated uncertainties are here 95\% statistical confidence bounds, in quantitative agreement with the expected number squeezing.

The measured squeezing presented here concerns the external degree
of freedom of the atoms in the condensate. It is comparable in amount to the latest measured squeezing on internal atomic states also obtained by
unitary evolution of a nonlinear Hamiltonian~\cite{chaudhury:2007le,fernholz:2008ez}. We show that the achieved entanglement can be directly used as a resource for quantum metrology with spatial atom interferometers. This is a major step towards measuring at the ultimate, Heisenberg, limit with a
large number of particles.

\begin{acknowledgments}
We gratefully acknowledge support from the DFG, GIF and EC (MIDAS STREP). J.E. acknowledges support from the EC Marie-Curie program. C.G. acknowledges support from the Landesgraduiertenf\"orderung Baden-W\"urttemberg.
\end{acknowledgments}


\pagebreak

\appendix

\renewcommand\thefigure{\Roman{figure}}
\setcounter{figure}{1}

\section{\sc \large  Supplementary Information} 

\section{Measuring Coherent Number Squeezing in a BEC: Experimental Techniques}

\subsection{Calculation of the number squeezing factor}
We measure the atom number per well using absorption imaging (see next section). For the considered pair of wells, the raw data consists of a set of atom numbers $n_a$ and $n_b$ obtained by repeating the experiment 25 to 40 times. Data shown in figure~3 and figure~4 represent an average over 4 such sets. For each set, we define $p = \langle n_a/(n_a + n_b)\rangle$ the probability for an atom to be in well $a$. We compute $\Delta n^2 = \langle \left[(n_a - n_b)/2 - (p-1/2)(n_a+n_b)\right]^2 \rangle$ to avoid taking into account fluctuations in the total atom number.  Photon shot-noise
contributes to the measured fluctuations of the atom number per well
by $\delta n_{a,{\rm psn}}$ and $\delta n_{b,{\rm psn}}$. These two
values are deduced from the light intensity on the absorption and on
the reference picture for each experimental realization. We subtract
this contribution and obtain the corrected number fluctuations
$\Delta n^2_{\rm corrected}  = \Delta n^2 - [1/4 + (p-1/2)^2] \langle
\delta n_{a,{\rm psn}}^2+\delta n_{b,{\rm psn}}^2 \rangle$. The subtracted photon shot noise corresponds to fluctuations of 10 to 12 atoms, which is comparable to the smallest fluctuations that we deduce after subtraction.
Finally the number squeezing factor is calculated by normalizing these
fluctuations by the expected value for a binomial distribution
$\xi_N^2 =  \Delta n^2_{\rm corrected}/[p(1-p) \langle n_a + n_b
\rangle]$.

\subsection{Atom number measurement and calibration}
\begin{figure}
\begin{center}
\includegraphics{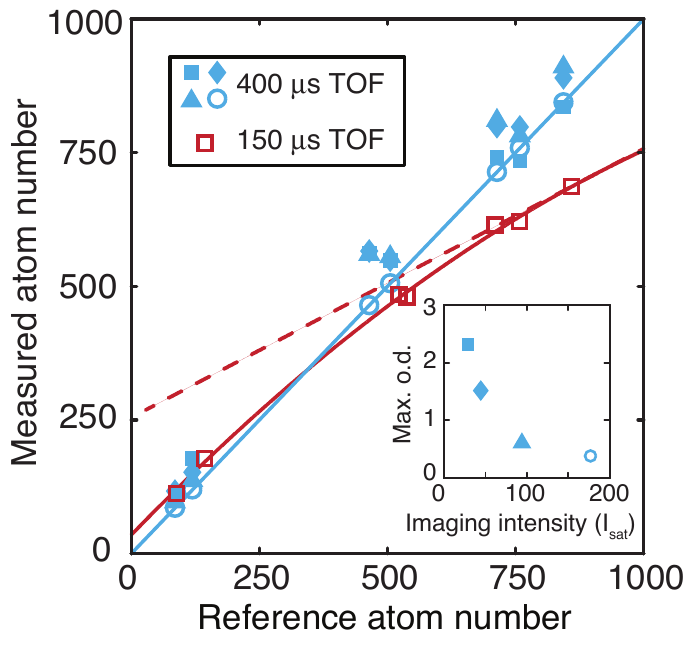}
\caption{Deducing the atom number from absorption images. Precise estimation of the atom number fluctuation requires a linear dependence of the measured atom number on the real atom number. The reference atom number defining the horizontal axis is measured in the linear regime, \emph{i.e.} at very high intensity compared to the saturation intensity $I_{\rm sat}$ (blue open circles). Blue solid symbols, corresponding to measurements taken after sufficient time of flight expansion (TOF) of the cloud, feature this linear behavior. We use this expansion time and an optical density around 1.5 for our measurements. Imaging atomic clouds smaller than the optical resolution at intermediate imaging intensities ($\approx 25\,I_{\rm sat}$) underestimates the mean atom number and even more its fluctuations. This is illustrated by the red data, where the dashed line is the tangent to the measured atom number for a real atom number of 750. The atom number is underestimated by only 15\% but the fluctuations by a factor of 2.}\label{fig.imaging}
\end{center}
\end{figure}

Our observations of number and coherent number squeezing rely on a precise measurement of the atom number in each well. Special care has to be taken to deduce the atomic density and thus the number of atoms from absorption images. In fluctuation measurements, an important issue is the linearity of the measurement method. While a small non-linearity can have a limited impact on the absolute value of the measured atom number, the deduced fluctuations can be dramatically wrong (see figure~\ref{fig.imaging}).

For signal to noise optimization, our measurements are performed at high imaging intensity compared to the saturation intensity but still in the optically dense regime. In this case, the atomic density is a non linear function of the absorption as detailed in~[G. Reinaudi \emph{et al.}, Opt. Lett. {\bf 32} 3143 (2007)]. In order to check that we correctly take into account this non-linearity, we first take pictures at high imaging intensity such that the cloud is optically thin and the optical transition is saturated for all atoms. In this regime, the deduced number depends linearly on the real atom number. We use this measurement as a reference atom number (horizontal axis in figure~\ref{fig.imaging}) and compare it to measurements taken at lower intensity and thus higher optical density. In this regime, it is essential to make sure that the optical density varies on a length bigger than the optical resolution of the imaging system. In figure~\ref{fig.imaging}, blue data show that after sufficient time of flight the nonlinearity is indeed correctly taken into account. In contrast, a too short expansion time leads to underestimation of the atom number and its fluctuation (red squares). 

\begin{figure}
\begin{center}
\includegraphics{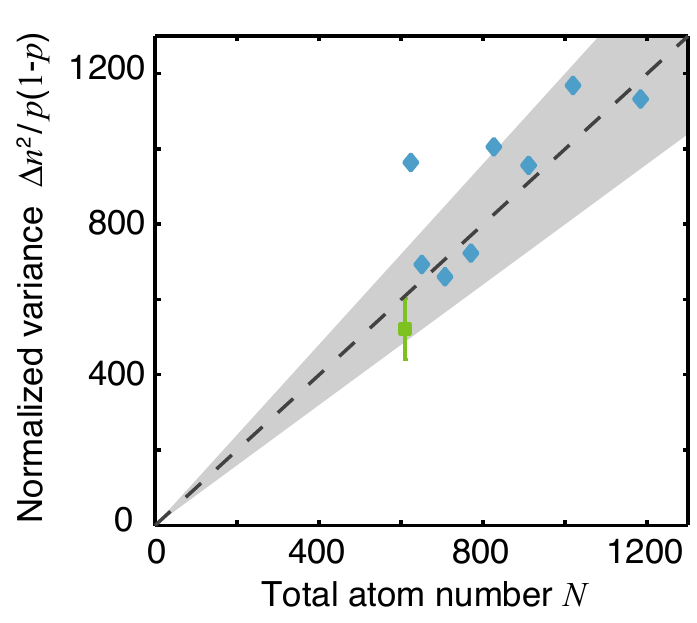}
\caption{Variance of the atom number difference as a function of the total atom number in a pair of wells. The data shown here correspond to two of the data sets plotted in the figure 4 of the article. For these samples, we expect the fluctuations of the atom number difference to be close to poissonian (dashed line) as explained in the text. The grey shaded area correspond to the $\pm 20$\% systematic error considered on the atom number that encompasses most of the data points. 
}\label{fig.variancevsmean}
\end{center}
\end{figure}

Once the linearity of the measurement is ensured, the conversion factor between the measured signal and the absolute atom number has still to be determined. This is done by comparing the observed density profiles of a Bose-Einstein condensate in a harmonic trap with numerical simulations of the three dimensional Gross-Pitaevskii equation for different atom numbers. An independent way to obtain the conversion factor are measurements in the high saturation regime, where only the transmission of the imaging system and the CCD sensitivity enter in the atom number determination. Measuring these parameters gives a calibration that agrees with the first method within 10\%. The given systematic error of 20\% on the total atom number and its fluctuations is thus a conservative upper bound. 

As explained in the article, we independently check the calibration by measuring the fluctuations of the atom number difference between two wells in a situation where they can be precisely calculated from first principles. For that, we utilize losses of particles as a way to restore close to poissonian fluctuations starting from a not well characterized initial number distribution. Evaporative cooling of the atomic cloud at a high lattice depth also leads to poissonian fluctuations. For this second method, the critical temperature to lattice depth ratio has to be well chosen in order to avoid any boson bunching effect that would cause super-poissonian fluctuations. In figure~\ref{fig.variancevsmean}, we plot the variance associated to these two different realizations of nearly poissonian samples as a function of the mean total atom number. We indeed observe that deviations from poissonian fluctuations indicated by the dashed line are small and stay within our systematic error estimation (grey shaded area). 

Starting from a slightly squeezed situation, the green point is measured after a long waiting time (10~s) during which losses significantly reduce the atom number (see article). The slightly remaining squeezing that we observe originates from the non-linear dependence of the three-body loss rate with the atom number. With the knowledge of the loss coefficients, this effect can be precisely calculated and match our data as shown in figure~4. 

The blue points are obtained by holding the atoms for different times after evaporative cooling in a deep lattice. In this case, the initial atom number is much lower than for the green point, leading to a negligible three-body loss rate. The initial nearly poissonian distribution remains poissonian under the action of one-body loss.

\subsection{Measuring the phase coherence}
The relative phase between two neighboring wells is deduced from local interference measurements. Due to the measurement process, high visibility fringes are present in each individual realization of the experiment even if the relative phase is not well defined. Fringes are observed after a short expansion in the harmonic trap in absence of the lattice potential ($2\,$ms) followed by a free expansion ($400\,\mu$s to $900\,\mu$s). In order to choose the proper timing, we image the cloud after different free expansion times and observe the formation of the interference pattern. For too short expansion times, clouds released from neighboring wells do not overlap which is easily seen in the images. In the case of a low lattice depth, all wells are in phase leading to a maximum of the interference pattern at the middle positions between the wells. We choose the timing such that this central maximum is clearly visible. 

The expansion velocity is dependent on the initial on-site interaction energy and thus on the atom number and on the barrier height. Since we measure the coherence for different final barrier heights, we always jump to a high barrier ($1650$~Hz) in 10~ms before releasing the atoms, such that the same expansion time can be used for all situations. For analysis, the observed fringe pattern is cut at the position of the wells such that in the resulting regions each sub pattern is dominated by the interference between the two next neighbor wells. The sub-patterns are Fourier transformed and the phase is measured at the frequency of the dominant component. We checked the validity of the method by simulating the expansion of the condensate using the three-dimensional Gross-Pitaevskii equation. 

The phase coherence $\cosphi$ between two neighboring wells is deduced by averaging the cosine of the different phases obtained from the sub-patterns corresponding to the wells of interest. The phase coherence would also correspond to the mean visibility of the fringes after ensemble averaging. The deduced coherence is a lower bound for the real coherence. Photon shot-noise in the absorption image as well as camera position instability lead to increased fluctuations of the fringe position and an underestimation of the coherence. All indicated numbers in this paper are not corrected for any of these noise sources. We consider them as systematic errors, as indicated in the text and in figure~2 by the horizontal width of the shaded areas.


\begin{thebibliography}{10}
\expandafter\ifx\csname url\endcsname\relax
  \def\url#1{\texttt{#1}}\fi
\expandafter\ifx\csname urlprefix\endcsname\relax\def\urlprefix{URL }\fi
\providecommand{\bibinfo}[2]{#2}
\providecommand{\eprint}[2][]{\url{#2}}

\bibitem{giovannetti:2004ix}
\bibinfo{author}{Giovannetti, V.}, \bibinfo{author}{Lloyd, S.} \&
  \bibinfo{author}{Maccone, L.}
\newblock \emph{\bibinfo{journal}{Science}} \textbf{\bibinfo{volume}{306}},
  \bibinfo{pages}{1330} (\bibinfo{year}{2004}).

\bibitem{santarelli:1999aa}
\bibinfo{author}{Santarelli, G.} \emph{et~al.}
\newblock \emph{\bibinfo{journal}{Phys. Rev. Lett.}}
  \textbf{\bibinfo{volume}{82}}, \bibinfo{pages}{4619--4622}
  (\bibinfo{year}{1999}).

\bibitem{goda:2008ix}
\bibinfo{author}{Goda, K.} \emph{et~al.}
\newblock \emph{\bibinfo{journal}{Nat. Phys.}} \textbf{\bibinfo{volume}{4}},
  \bibinfo{pages}{472--476} (\bibinfo{year}{2008}).

\bibitem{arcizet:2006eu}
\bibinfo{author}{Arcizet, O.} \emph{et~al.}
\newblock \emph{\bibinfo{journal}{Phys. Rev. Lett.}}
  \textbf{\bibinfo{volume}{97}}, \bibinfo{pages}{133601}
  (\bibinfo{year}{2006}).

\bibitem{kitagawa:1993wd}
\bibinfo{author}{Kitagawa, M.} \& \bibinfo{author}{Ueda, M.}
\newblock \emph{\bibinfo{journal}{Phys. Rev. A}} \textbf{\bibinfo{volume}{47}},
  \bibinfo{pages}{5138--5143} (\bibinfo{year}{1993}).

\bibitem{wineland:1994rr}
\bibinfo{author}{Wineland, D.~J.}, \bibinfo{author}{Bollinger, J.~J.},
  \bibinfo{author}{Itano, W.~M.} \& \bibinfo{author}{Heinzen, D.~J.}
\newblock \emph{\bibinfo{journal}{Phys. Rev. A}} \textbf{\bibinfo{volume}{50}},
  \bibinfo{pages}{67--88} (\bibinfo{year}{1994}).

\bibitem{sorensen:2001wd}
\bibinfo{author}{Sorensen, A.}, \bibinfo{author}{Duan, L.},
  \bibinfo{author}{Cirac, J.} \& \bibinfo{author}{Zoller, P.}
\newblock \emph{\bibinfo{journal}{Nature}} \textbf{\bibinfo{volume}{409}},
  \bibinfo{pages}{63--6} (\bibinfo{year}{2001}).

\bibitem{wang:2003eu}
\bibinfo{author}{Wang, X.} \& \bibinfo{author}{Sanders, B.~C.}
\newblock \emph{\bibinfo{journal}{Phys. Rev. A}} \textbf{\bibinfo{volume}{68}},
  \bibinfo{pages}{012101} (\bibinfo{year}{2003}).

\bibitem{korbicz:2005oq}
\bibinfo{author}{Korbicz, J.~K.}, \bibinfo{author}{Cirac, J.~I.} \&
  \bibinfo{author}{Lewenstein, M.}
\newblock \emph{\bibinfo{journal}{Phys. Rev. Lett.}}
  \textbf{\bibinfo{volume}{95}}, \bibinfo{pages}{120502}
  (\bibinfo{year}{2005}).

\bibitem{hald:1999lh}
\bibinfo{author}{Hald, J.}, \bibinfo{author}{S\o{}rensen, J.~L.},
  \bibinfo{author}{Schori, C.} \& \bibinfo{author}{Polzik, E.~S.}
\newblock \emph{\bibinfo{journal}{Phys. Rev. Lett.}}
  \textbf{\bibinfo{volume}{83}}, \bibinfo{pages}{1319--1322}
  (\bibinfo{year}{1999}).

\bibitem{kuzmich:2000bv}
\bibinfo{author}{Kuzmich, A.}, \bibinfo{author}{Mandel, L.} \&
  \bibinfo{author}{Bigelow, N.~P.}
\newblock \emph{\bibinfo{journal}{Phys. Rev. Lett.}}
  \textbf{\bibinfo{volume}{85}}, \bibinfo{pages}{1594--1597}
  (\bibinfo{year}{2000}).

\bibitem{julsgaard:2001oz}
\bibinfo{author}{Julsgaard, B.}, \bibinfo{author}{Kozhekin, A.} \&
  \bibinfo{author}{Polzik, E.~S.}
\newblock \emph{\bibinfo{journal}{Nature}} \textbf{\bibinfo{volume}{413}},
  \bibinfo{pages}{400--403} (\bibinfo{year}{2001}).

\bibitem{chaudhury:2007le}
\bibinfo{author}{Chaudhury, S.} \emph{et~al.}
\newblock \emph{\bibinfo{journal}{Phys. Rev. Lett.}}
  \textbf{\bibinfo{volume}{99}}, \bibinfo{pages}{163002}
  (\bibinfo{year}{2007}).

\bibitem{fernholz:2008ez}
\bibinfo{author}{Fernholz, T.} \emph{et~al.}
 \newblock \emph{\bibinfo{journal}{Phys. Rev. Lett.}}
  \textbf{\bibinfo{volume}{101}}, \bibinfo{pages}{073601}
  (\bibinfo{year}{2008}).

\bibitem{meyer:2001hc}
\bibinfo{author}{Meyer, V.} \emph{et~al.}
\newblock \emph{\bibinfo{journal}{Phys. Rev. Lett.}}
  \textbf{\bibinfo{volume}{86}}, \bibinfo{pages}{5870--5873}
  (\bibinfo{year}{2001}).

\bibitem{leibfried:2004hs}
\bibinfo{author}{Leibfried, D.} \emph{et~al.}
\newblock \emph{\bibinfo{journal}{Science}} \textbf{\bibinfo{volume}{304}},
  \bibinfo{pages}{1476--1478} (\bibinfo{year}{2004}).

\bibitem{roos:2004dp}
\bibinfo{author}{Roos, C.~F.} \emph{et~al.}
\newblock \emph{\bibinfo{journal}{Science}} \textbf{\bibinfo{volume}{304}},
  \bibinfo{pages}{1478--1480} (\bibinfo{year}{2004}).

\bibitem{orzel:2001aa}
\bibinfo{author}{Orzel, C.}, \bibinfo{author}{Tuchman, A.},
  \bibinfo{author}{Fenselau, M.}, \bibinfo{author}{Yasuda, M.} \&
  \bibinfo{author}{Kasevich, M.}
\newblock \emph{\bibinfo{journal}{Science}} \textbf{\bibinfo{volume}{291}},
  \bibinfo{pages}{2386} (\bibinfo{year}{2001}).

\bibitem{greiner:2002aa}
\bibinfo{author}{Greiner, M.}, \bibinfo{author}{Mandel, O.},
  \bibinfo{author}{Hansch, T.} \& \bibinfo{author}{Bloch, I.}
\newblock \emph{\bibinfo{journal}{Nature}} \textbf{\bibinfo{volume}{419}},
  \bibinfo{pages}{51--4} (\bibinfo{year}{2002}).

\bibitem{gerbier:2006cr}
\bibinfo{author}{Gerbier, F.}, \bibinfo{author}{F\"{o}lling, S.},
  \bibinfo{author}{Widera, A.}, \bibinfo{author}{Mandel, O.} \&
  \bibinfo{author}{Bloch, I.}
\newblock \emph{\bibinfo{journal}{Phys. Rev. Lett.}}
  \textbf{\bibinfo{volume}{96}}, \bibinfo{pages}{090401}
  (\bibinfo{year}{2006}).

\bibitem{sebby-strabley:2007nx}
\bibinfo{author}{Sebby-Strabley, J.} \emph{et~al.}
\newblock \emph{\bibinfo{journal}{Phys. Rev. Lett.}}
  \textbf{\bibinfo{volume}{98}}, \bibinfo{pages}{200405}
  (\bibinfo{year}{2007}).

\bibitem{jo:2007hl}
\bibinfo{author}{Jo, G.-B.} \emph{et~al.}
\newblock \emph{\bibinfo{journal}{Phys. Rev. Lett.}}
  \textbf{\bibinfo{volume}{98}}, \bibinfo{pages}{030407}
  (\bibinfo{year}{2007}).

\bibitem{li:2007jx}
\bibinfo{author}{Li, W.}, \bibinfo{author}{Tuchman, A.~K.},
  \bibinfo{author}{Chien, H.-C.} \& \bibinfo{author}{Kasevich, M.~A.}
\newblock \emph{\bibinfo{journal}{Phys. Rev. Lett.}}
  \textbf{\bibinfo{volume}{98}}, \bibinfo{pages}{040402}
  (\bibinfo{year}{2007}).

\bibitem{sorensen:2001ax}
\bibinfo{author}{S\o{}rensen, A.~S.} \& \bibinfo{author}{M\o{}lmer, K.}
\newblock \emph{\bibinfo{journal}{Phys. Rev. Lett.}}
  \textbf{\bibinfo{volume}{86}}, \bibinfo{pages}{4431--4434}
  (\bibinfo{year}{2001}).

\bibitem{korbicz:2006kb}
\bibinfo{author}{Korbicz, J.~K.} \emph{et~al.}
\newblock \emph{\bibinfo{journal}{Phys. Rev. A}} \textbf{\bibinfo{volume}{74}}, \bibinfo{pages}{052319}
  (\bibinfo{year}{2006}).

\bibitem{javanainen:1999ab}
\bibinfo{author}{Javanainen, J.}
\newblock \emph{\bibinfo{journal}{Phys. Rev. A}}
  \textbf{\bibinfo{volume}{60}}, \bibinfo{pages}{4902--4909}
  (\bibinfo{year}{1999}).

\bibitem{gati:2006ab}
\bibinfo{author}{Gati, R.}, \bibinfo{author}{Hemmerling, B.},
  \bibinfo{author}{Folling, J.}, \bibinfo{author}{Albiez, M.} \&
  \bibinfo{author}{Oberthaler, M.~K.}
\newblock \emph{\bibinfo{journal}{Phys. Rev. Lett.}}
  \textbf{\bibinfo{volume}{96}}, \bibinfo{pages}{130404}
  (\bibinfo{year}{2006}).

\bibitem{reinaudi:2007nx}
\bibinfo{author}{Reinaudi, G.}, \bibinfo{author}{Lahaye, T.},
  \bibinfo{author}{Wang, Z.} \& \bibinfo{author}{Gu{\'e}ry-Odelin, D.}
\newblock \emph{\bibinfo{journal}{Optics Letters}}
  \textbf{\bibinfo{volume}{32}}, \bibinfo{pages}{3143--3145}
  (\bibinfo{year}{2007}).

\bibitem{esteve:2005ab}
\bibinfo{author}{Est{\`e}ve, J.} \emph{et~al.}
\newblock \emph{\bibinfo{journal}{Phys. Rev. Lett.}}
  \textbf{\bibinfo{volume}{96}}, \bibinfo{pages}{130403}
  (\bibinfo{year}{2005}).

\end{thebibliography}
\end{document}